\documentclass[]{raa}            
\usepackage{graphicx,times}
\usepackage{natbib}

\providecommand{\bm}[1]{\mbox{\boldmath$#1$\unboldmath}}

\begin{document}

   \title{Long-term Continuous Energy Injection in the Afterglow of GRB 060729
}

 \volnopage{ {\bf 2009} Vol.\ {\bf 9} No. {\bf XX}, 000--000}
   \setcounter{page}{1}

   \author{M. Xu
      \inst{1}
   \and Y.-F. Huang
      \inst{1}
   \and T. Lu
      \inst{2}
   }

  \institute{Department of Astronomy, Nanjing University, Nanjing
210093, China; {\it hyf@nju.edu.cn}\\
  \and
    Purple Mountain Observatory, Chinese Academy of Sciences, Nanjing 210008,
    China\\
        \vs \no
   {\small Received [year] [month] [day]; accepted [year] [month] [day] }
}

\abstract{ A long plateau phase and an amazing brightness have been
observed in the X-ray afterglow of GRB 060729. This peculiar light
curve is likely due to long-term energy injection in external shock.
Here we present a detailed numerical study on the energy injection
process of magnetic dipole radiation from a strongly magnetized
millisecond pulsar and model the multi-band afterglow observations.
It is found that this model can successfully explain the long
plateaus in the observed X-ray and optical afterglow light curves.
The sharp break following the plateaus should be due to the rapid
decline of the emission power of the central pulsar. At an even late
time ($\sim 5\times 10^{6}s$), an obvious jet break appears, which
implies a relatively large half opening angle of $\theta \sim 0.3$
for the GRB ejecta. Due to the energy injection, the Lorentz factor
of the outflow is still larger than two $10^{7}$s post the GRB
trigger, making the X-ray afterglow of this burst detectable by
Chandra even 642 days after the burst. \keywords{ gamma rays: bursts
-ISM: jets and outflows }
 }

   \authorrunning{M. Xu, Y.-F. Huang \& T. Lu }            
   \titlerunning{Long-term Energy Injection in GRB 060729 }  
   \maketitle


%
%
\section{Introduction}           
\label{sect:intro}

GRB 970228 is the first gamma-ray burst (GRB) with an X-ray
afterglow detected (Costa et al. 1997). Optical (van Paradijs et al.
1997) and radio afterglow (Frail et al. 1997) has also been
unprecedently detected from this event. The relativistic internal
and external shock model is the most successful model to explain
these violent events ( Rees \& M\'{e}sz\'{a}ros 1994; Piran 1999;
Zhang 2007). It is also widely believed that long GRBs should be due
to the collapse of massive stars (Woosley 1993; Paczy\'{e}ski 1998;
MacFadyen \& Woosley 1999), and short GRBs should be connected with
the coalescence of two compact objects (Eichler et al. 1989; Narayan
et al. 1992; Gehrels et al. 2005; Nakar 2007).

The X-ray telescope (XRT) on board Swift reveals that the X-ray
afterglows of GRBs generally show a canonical behavior, with five
components in the observed X-ray afterglow light curves, i.e., steep
decay phase, shallow decay phase, normal decay phase, post jet break
phase and X-ray flares (Zhang et al. 2006; Nousek et al. 2006). The
conventional models for shallow decay phase are energy injection
from strongly magnetized millisecond pulsar (Dai \& Lu 1998; Zhang
\& M\'{e}sz\'{a}ros 2001; Liang et al. 2007; Lyons et al. 2009) or
from ejecta with a highly dispersed Lorentz factor distribution
(Rees \& M\'{e}sz\'{a}ros 1998; Sari \& M\'{e}sz\'{a}ros 2000).

At $19:12:29$ UT of July 29, 2006, GRB 060729 triggered the $Swift$
Burst Alert Telescope (BAT) and was quickly located
(Grupe et al. 2006). This event has a duration of
$T_{90}=116\pm10s$ (Parsons et al. 2006) and a redshift of $z=0.54$
(Thoene et al. 2006). The isotropic energy release in the rest-frame
in $1keV-10MeV$ band was $E_{iso}=1.6\times10^{52}ergs$ for a
standard cosmology model with $\Omega_{M}=0.27$,
$\Omega_{\Lambda}=0.73$ and a Hubble constant of $H_{0}=71 km\cdot
s^{-1}\cdot Mpc^{-1}$.

One of the distinguished properties of GRB 060729 is that it has a
long flat phase in the X-ray afterglow light curve (Grupe et al.
2007). Another prominent character of GRB 060729 is its brightness.
It can be observed by $Chandra$ even $642$ days after the burst
trigger (Grupe et al. 2009). Grupe et al. (2009) compared
the X-ray afterglow of GRB 060729 with other bright X-ray afterglows
and concluded that GRB 060729 was an exceptionally long-lasting
event. Actually, the brightness of the X-ray afterglow of GRB 060729
is not extraordinary at early time ($t<30000s$), but it becomes the
brightest one among all GRBs after $30000s$ since the trigger.

In view of the long plateau phase ($500s-30000s$) and the late time
($>30000s$) brightness of GRB 060729, a strong and long-term
continuous energy injection is implied (Liang et al. 2007; Grupe et
al. 2007, 2009). Grupe et al. (2007) presented an extensive study on
this peculiar event and made a detailed analysis on the pulsar-type
energy injection for this plateau. But at that time there was only
125 days of data and the jet break still did not appear.  In this
paper, we use the energy injection model that involves the dipole
radiation from a strongly magnetized millisecond pulsar to explain
the special behavior of the multi-band afterglow of GRB 060729. The
new data observed by $Chandra$ (Grupe et al. 2009) will be
incorporated. We detailedly calculate the X-ray and optical (U-band,
B-band and V-band) afterglow light curves, and compare them with the
observations. In Section 2, we briefly describe the energy injection
model. In Section 3 we present our detailed numerical results.
Finally, Section 4 is our conclusions and discussion.

\section{Energy Injection from a Strongly Magnetized Millisecond Pulsar}
\label{sect:EI}

Due to the strong magnetic field and rapid rotation, a new born
millisecond pulsar will radiate a huge amount of energy through magnetic dipole
emission. This energy can be comparable to or even larger than the
initial energy of the main GRB. Detailed discussions on this process
have been given by Dai \& Lu (1998) and Zhang \& M\'{e}sz\'{a}ros
(2001).

Through magnetic dipole radiation, the new born pulsar in the center
of the GRB fireball will lose its rotational energy. The radiation power
evolves with time as
\begin{equation}
 L=L_{0}(1+ \frac{t}{T})^{-2} ,
 \label{L}
\end{equation}
where $L_{0}$ is the initial luminosity, i.e., the radiation power at
the time of $t=0$. $T$ is the characteristic spin-down timescale.

The initial luminosity depends on the parameters of the pulsar as
\begin{equation}
 L_{0}=4.0\times10^{47}ergs\cdot s^{-1}
 (B^{2}_{\bot,14}P^{-4}_{-3}R^{6}_{6}),
 \label{L0}
\end{equation}
where $B_{\bot,14}=B_{s}sin\vartheta/10^{14}G$, $B_{s}$ is the strength
of the dipole magnetic field at the surface of the pulsar, $\vartheta$
is the angle between the rotation axis and the magnetic axis,
$P_{-3}$ is the pulsar period in units of $10^{-3}s$, and $R_{6}$ is
the radius of the pulsar in units of $10^{6}cm$.

The characteristic spin-down timescale of the pulsar can be calculated
from
\begin{equation}
 T=5.0\times10^{4}s(B^{-2}_{\bot,14}P^{2}_{-3}R^{-6}_{6}I_{45}),
 \label{T}
\end{equation}
where $I_{45}$ is the moment of inertia of the pulsar in units of
$10^{45}g\cdot cm^{2}$.

The total energy of the magnetic dipole radiation can be derived
by integrating the emission power from $t=0$ to
$t\rightarrow\infty$
\begin{equation}
 E_{total}=\int_{0}^{\infty}Ldt=\int_{0}^{\infty}[L_{0}(1+
 \frac{t}{T})^{-2}]dt=L_{0}T.
 \label{Etot}
\end{equation}

\section{Numerical Calculation and Results}
\label{sect:cal}

A convenient method to describe the dynamics and radiation
processes of GRB afterglows has been proposed by Huang et
al. (2000). It is appropriate for both radiative and adiabatic
blastwaves, and in both the ultra-relativistic and the
non-relativistic phases (Huang et al. 1999). Here we modify
their method accordingly so that
it can be applicable to the energy injection scenario.

\subsection{Dynamics}

The overall dynamical evolution of GRB afterglows has been described by Huang et
al. (1999, 2000). When the energy injection from a strongly
magnetized millisecond pulsar is included, the deceleration of the external shock
is mainly characterized by the
following equation
\begin{equation}
 \frac{d \gamma}{d m} = \frac{-(\gamma^2 -
  1)+d(Lt)/d(mc^{2})}
  {M_{\rm ej} + \epsilon m + 2 ( 1 - \epsilon) \gamma m},
 \label{dgdm}
\end{equation}
where $\gamma$ is the bulk Lorentz factor of the shocked medium, $m$
is the swept-up mass, $M_{ej}$ is the initial ejecta mass, and
$\epsilon$ is the radiation efficiency.

For simplicity, here we only consider the synchrotron emission from
shock-accelerated electrons. To get the observed afterglow flux,
we need to integrate the emission power over the equal arrival time
surface determined by
\begin{equation}
 \int \frac{1 - \beta \cos \Theta}{\beta c} dR \equiv t,
 \label{eqt}
\end{equation}
within the jet boundaries, where
$\beta=\sqrt{\gamma^{2}-1}/\gamma$ and $\Theta$ is the angle between
the velocity of emitting material and the line of sight.

\subsection{Numerical Results}
Inserting Eq. (1) into Eq. (5), we can conveniently calculate
the evolution of the external shock subject to the energy injection
from a strongly magnetized millisecond pulsar. In this section,
we assume that the circum-burst medium is homogeneous.
We calculate the overall dynamical evolution of a uniform jet
to educe the X-ray and optical afterglow light curves, and try to
give the best fit to the observations of GRB 060729.

To get the best fit, we find that we need to set the parameters
of the central pulsar as follows. The radius is $R_{6}=1$.
The rotation period is $P_{-3}=1.49$. The magnetic field
is $B_{\bot,14}=2.72$. The moment of inertia is
taken as $I_{45}\sim 2$, which is still typical for
neutron stars (Datta 1988; Weber \& Glendenning 1993).
Then, according to equations $(2)$ and $(3)$, the initial emission
power and the spin-down timescale of the center pulsar are
$L_{0}=6.0\times 10^{47}ergs\cdot s^{-1}$, and $T=30000s$
respectively. So the energy injection power is $L=6.0\times
10^{47}ergs\cdot s^{-1}(1+ \frac{t}{30000 {\rm s}})^{-2}$.

In our calculations, we use the following parameters for
the external shock of GRB 060729:
initial energy per solid angle $E_{0} =1.6\times 10^{52}/4\pi$
ergs, the initial Lorentz factor $\gamma_0 = 200$, the $ISM$ number
density $n = 0.2 cm^{-3}$, the power-law index of the energy
distribution of electrons $p = 2.48$, the luminosity distance $D_{L}
= 3.12$ Gpc, the electron energy fraction $\epsilon_{e} = 0.15$, the
magnetic energy fraction $\epsilon_{B} = 0.0002$, the half
opening angle of the jet $\theta= 0.3$, and the observing angle $\theta_{obs} = 0$.
Here the observing angle is defined as the angle between the line
of sight and the jet axis.

Using the above parameter set, we can give a satisfactory fit to the
multiband afterglows of GRB 060729. In Fig.~1, we first show the
evolution of the Lorentz factor under the energy injection from a
strongly magnetized millisecond pulsar. We see that due to the
continuous energy injection, the Lorentz factor of the outflow is
still larger than 2 after $10^{7}$s. It means that the afterglow
could be very bright even at very late stages.

Fig.2 illustrates the observed X-ray (0.3-10 keV) afterglow light
curve of GRB 060729 and our best fit. We can see that the observed
X-ray afterglow light curve is fitted very well. Especially, the
observed long plateau ($~500s-30000s$) is explained satisfactorily.
This long flat phase is resulted from the long-term continuous
energy injection from the magnetic dipole radiation of the strongly
magnetized millisecond pulsar. After $30000s$, the flat phase comes
to the end and a break is seen in the light curve. The reason is
that the pulsar has consumed most of its rotation energy on the
spin-down timescale (T=30000s in our model), so that the power of
energy injection decreases sharply at that time. An obvious jet
break is presented at $t_{j}\sim 5\times 10^{6}s$. To produce such a
late jet break, we find that the half opening angle of the jet
should be $\theta=0.3$, which is relatively large among known GRBs.

Fig.3 illustrates our fit to the observed optical afterglows of GRB
060729 by using the same parameters as in Figs.~1 and 2. All the
data points are taken from Grupe et al. (2007). We see that the
observed optical afterglow can also be satisfactorily explained.

\section{Conclusions and Discussion}
\label{sect:cd}

We have shown that the observed special behavior of the afterglow of
GRB 060729 can be well explained by using the energy injection
model. Our study indicates that the central engine should be a
strongly magnetized millisecond pulsar, which continuously supplies
energy to the GRB ejecta via magnetic dipole radiation on a
timescale of about 30000 s. The observed multi-band afterglow light
curves can be reproduced satisfactoriely by this model. According to
our calculations, the duration of the plateau phase in the afterglow
light curve should correspond to the spin-down timescale of the
pulsar ($T=30000s$). To further explain the observed jet break at
$t_{j}\sim 5\times 10^{6}s$, we need a relatively large jet opening
angle of $\theta=0.3$.

From Equation (4), we can derive the total injected energy
as $E_{total}=L_{0}T=1.8\times 10^{52}ergs$. This energy is comparable
to the initial isotropic energy release in the main burst phase
($E_{iso}=1.6\times 10^{52}ergs$). The long-term continuous energy
injection makes GRB 060729 the brightest burst in X-ray band at late
stages. In fact, the X-ray afterglow can be
observed even $642$ days after the trigger (Grupe et al. 2009).

In optical bands, the afterglow light curves show some similar
properties as in the X-ray band. For example, a flat stage is
presented in the optical light curves. Generally, our model can give
a satisfactory explanation to the optical afterglow. The time span
of optical observations is very limited. We do not have optical data
for $t > 10^6$ s, so that the jet break is still not observed in
optical band. However, note that the extensive analysises with both
the X-ray and optical data in recent years show that some of the
jet-like breaks in the afterglow light curves are chromatic
(Panaitescu et al. 2006; Liang et al. 2008). The nature of these
breaks is then highly debatable. Thus a long term monitoring of the
multi-band afterglows is definitely necessary. Also note that the
early UV-optical afterglow light curves of GRB 060729 show
significant variations. This feature, however, is not seen in the
X-ray light curve. It indicates that other regions may also
contribute to the optical emission in this event.

In our current study, we have assumed that the energy injection is
isotropic. Magnetic dipole radiation actually should be anisotropic.
However, this kind of anisotropy is not significant and would not
affect the final results seriously.

According to our numerical results, the Lorentz factor of the jet is
still larger than 2 after $10^{7}$s. This is due to the continuous
and long-term energy injection. As a result, the time that the
afterglow of GRB 060729 enters the Newtonian phase is significantly
delayed.

The energy injection models were used to explain the afterglows of
some GRBs, such as GRB 010222 (Bj\"{o}rnsson et al. 2002), GRB
021004 (Bj\"{o}rnsson et al. 2004), GRB 030329 (Huang et al. 2006)
and GRB 051221A (Fan \& Xu 2006) etc. The explanation of the
afterglow from the short GRB 051221A also needs some kind of energy
injection from a magnetar (Fan \& Xu 2006). However, we note that
the physical origin of the shallow decay segment is still highly
debating (Zhang 2007). Generally speaking, while the achromatic
breaks in both the X-ray and the optical bands can be explained with
conventional energy injection models, the chromatic breaks of this
segment observed in many events strongly challenge these models
(Liang et al. 2007). Alternative models that go beyond the
conventional ones were proposed (see Zhang 2007 for review). It is
interesting that a small fraction of XRT lightcurves show as a
single power-law without canonical feature (Liang et al. 2009). It
was also argued that the apparent difference of the canonical and
single power-law XRT lightcurves may be due to the improper zero
time effect on the canonical XRT lightcurves (Yamazaki 2009; Liang
et al. 2009).

\normalem
\begin{acknowledgements}
We thank the anonymous referee for helpful suggestions. This work
was supported by the National Natural Science Foundation of China
(Grant No. 10625313 \& 10473023) and the National Basic Research
Program of China (973 Program, grant 2009CB824800).
\end{acknowledgements}

\begin{figure}[h!!!]
   \centering
   \includegraphics[width=10.0cm, angle=0]{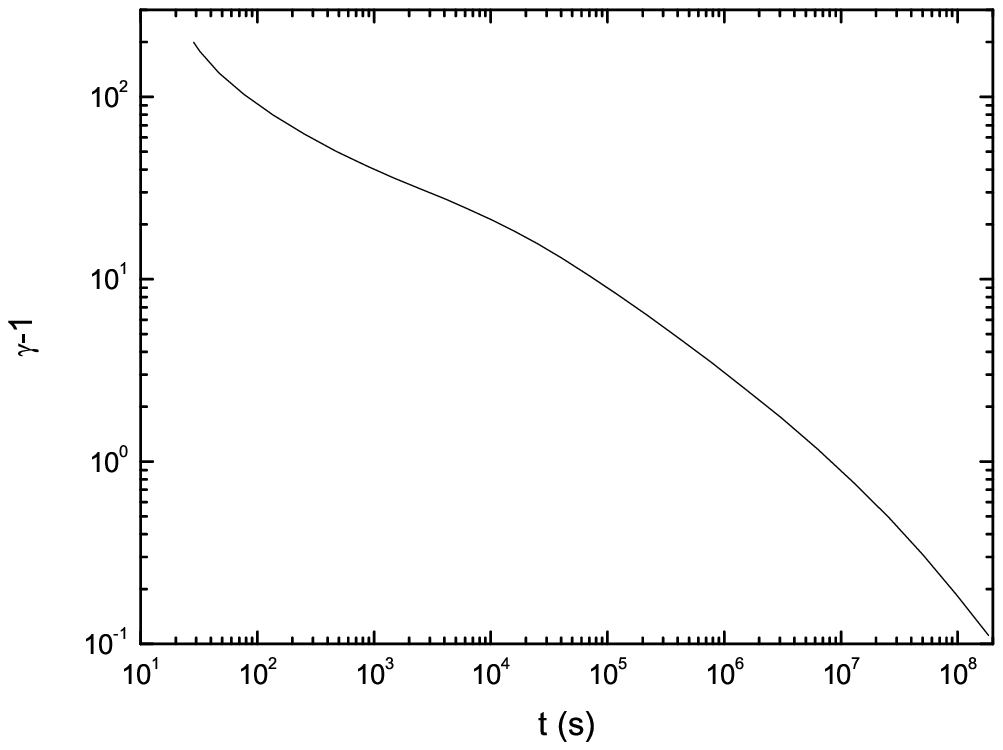}
   \caption{ Evolution of the bulk Lorentz factor of a jet with long-term energy
    injection from a strongly magnetized millisecond pulsar. The parameters used in
    this calculation have been given in Section 3.2 .}
   \label{Fig1}
\end{figure}

\begin{figure}
   \centering
   \includegraphics[width=10.0cm, angle=0]{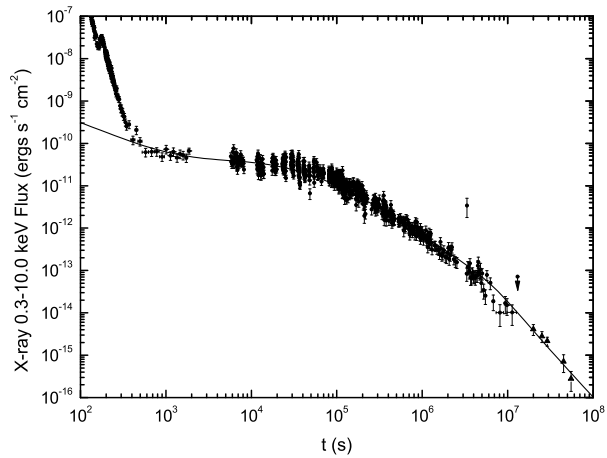}
   \caption{ Observed X-ray afterglow light curve of GRB 060729 and our best fit
   by using the energy injection model. The square points are observed data from
   \emph{Swift} and the triangle
   points are observed data from \emph{Chandra} (Grupe et al. 2009).
   The tail emission in the very
   early phase ($t < 400$ s) is not considered in our fit. }
   \label{Fig2}
\end{figure}

\begin{figure}
  \centering
  \includegraphics[width=10.0cm, angle=0]{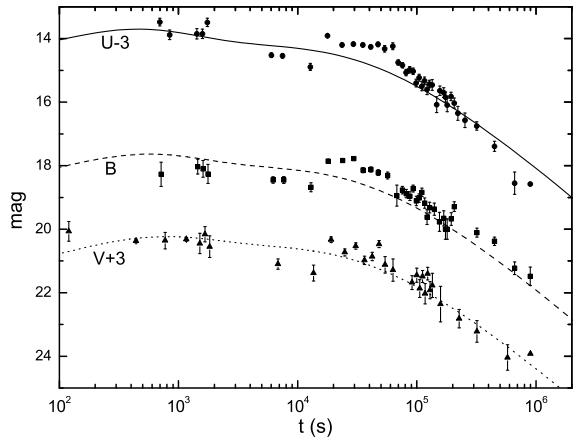}
  \caption{Observed multi-band optical afterglow light curves of GRB
   060729 and our best fit by using the energy injection model.
   Observed data points are taken from Grupe et al. (2007).
   The solid, dashed and dotted lines are our fit to
   the observed light curves in the three bands, respectively.
   Note that the U and V-band light curves have been shifted
   by 3 magnitudes for clarity.
    }
  \label{Fig2}
\end{figure}

\label{lastpage}

\end{document}